# Balloon-to-Balloon AdHoc Wireless Network Connectivity: Google Project Loon

Aishwarya Srinivasan[1]
Data Scientist, Google Cloud
aishgrt@gmail.com

***Abstract:*** *Project Loon is a Google initiated research project from the Google X Lab. The project focuses on providing remote internet access and network connectivity. The connectivity is established in vertical and horizontal space; vertical connectivity between Google Access Point (GAP) and the balloons, and between balloons and antennas installed at land; horizontal connectivity is between the balloons. This research focuses on the connectivity between the balloons in a mesh network. The proposal focuses on implementing graphical methods like convex hull with adhoc communication protocols. The proposed protocol includes content-based multicasting using angular sector division rather than grids, along with dynamic core-based mesh protocol defining certain core active nodes and passive nodes forming the convex hull. The transmission (multicasting and broadcasting) between the nodes will be evaluated using the link probability defining the probability of the link between two nodes failing. Based on the link probability and node features, best path between transmitting and receiver nodes will be evaluated.*

## I. INTRODUCTION

The Project Loon works with a floating parachute type balloon structure which is mobile and acts as a transmitter of network connectivity. The Loon balloons float in the stratosphere similar to the weather balloons which collect atmospheric information using sensors. But, unlike weather balloons, Loon balloons are super pressure balloons that can stay for more than 100 days.

The Loon balloons are so created that they float in the atmosphere carrying some sophisticated hardware and software components that are meant to provide internet access to the people who are willing to connect to it. The range of each balloon is limited, wherein they can provide network connectivity. People connect to the network provided by each balloon via an antenna which is installed at their offices or houses. The signal bounces from balloon to balloon, then to global internet and back to earth. The connectivity among the balloons is made through a mesh network, to increase the robustness for high-speed internet. Each balloon can provide connectivity to the ground area for about 40 km in diameter and with speed comparable to 3G. For the setup of the Loon balloons, three segments of communication can be specified; balloon to Google Access point (GAP), balloon to balloon and balloon to user communication.

Protocols may be so designed that a higher bandwidth signal is diverted to the balloons providing connectivity to an area with higher demand. The Google Loon balloons are currently using ISM bands of 2.4 to 5.8 GHz. The balloon to balloon communication is enabled by special radio frequency technology.

Project Loon balloons float in the stratosphere, twice as high as commercial airplanes and the weather. In the stratosphere, there are many layers of wind, and each layer of wind varies in direction and speed. Loon balloons go where they're needed by rising or descending into a layer of wind blowing in the desired direction of travel. People can connect to the balloon network using a special Internet antenna attached to their building. The signal bounces from this antenna up to the balloon network and then down to the global Internet on Earth.

## II. WORKING AND METHODOLOGY

With the advent of sophisticated wireless technology and the Internet of Things (IoT) concept, wireless communication and sensor networks are paving the way for reshaping the method of internetwork communication. The Internet of Things concept highlights the communication between the passive objects, through the wireless network. Whereas, the latest technological development which is being introduced is the Internet of Everything (IoE), which focuses on not just the communication between the passive objects, but also the communication of these objects with the users. The information transfer which has been made feasible between the nodes can be modified to make their utility with the human needs.

Project Loon uses software algorithms to determine where its balloons need to go, and then moves each one into a layer of wind blowing in the right direction. By moving with the wind, the balloons can be arranged to form one large communications network. Project Loon began with a pilot test in June 2013, when 30 balloons were launched from New Zealand's South Island and beamed the Internet to local pilot testers. The test has since expanded to include a greater number of people over a wider area.

Google's Project Loon focuses on the purpose of providing a network to everyone. In many of the remote areas where the infrastructure is not well established to have full-fledged internet connectivity can be benefitted by the wireless network-based balloons that act as network connection transmitters. In situations where any place has been affected by a natural disaster which led to a huge loss of infrastructure, due to which the connectivity in that region, be it internet connectivity or the mobile phone networks can be pulled over by these remote balloons which serve the purpose.

## III. TECHNOLOGY BREAKTHROUGHS

Project Loon incorporates technology breakthroughs; namely fleet planning and steering. In fleet planning, the Google Access Point (GAP) plans the paths of tens of thousands of balloons, along with taking into consideration the energy cost constraints. The movement of the balloons can be controlled directly from the GAP which has been put as the steering feature.

Much development has been made in the field of putting up Unmanned Aerial Vehicles (UAVs), which have radio equipment attached to them along with sophisticated computational instruments. The main limitation in the Aerial Vehicles was that the cost of making them and maintaining them was very high.

Google reports produce that Project Loon consists of two types of network communication:
1. Vertical Network: Between balloon & users

2. Mesh Network: Between balloons

The network communication system includes balloon to GAP, balloon to balloon, and balloon to user communication. The balloon to user communication is carried out with the IEEE 802.11 b, g and the balloon to balloon communication is supported with IEEE 802.11 j.

## IV. INTERNETWORKING AND TOPOLOGY

The wireless mesh network between the balloons is made up of the auto-configuration functionality, which operates on the electromagnetic field power density. It can be better explained as a balloon tries to find one of its neighbor balloons which have the strongest power density and establishes a connection with it. Then, the mesh network has a minimum spanning tree.

In the proposal, each balloon is associated with the following variables: load/ max load, Bandwidth, position, Signal strength, Neighboring nodes, and reachability. These are some of the factors which are generally considered, apart from which the proposal takes into consideration the altitude and the polar coordinates along with the radial distance.

In the Loon Project, there is no fixed infrastructure apart from the Google Access Point or the base station which controls the balloons. The mesh network to be formed between the balloons needs to be dynamic and adaptive.

As the linking status between nodes is unstable, and nodes themselves are also not fixed, forming a mesh network between the balloons becomes challenging. Hence, the concept of random graphs can be applied to tackle this issue. The phase transition probability may be considered before the formation of each of the links between the balloons.

As mentioned earlier, we would now discuss the variables or factors related to each balloon. The load is the number of users a balloon is serving. Along with the load, the demand of each user connected to the balloon may also vary. It is possible that a balloon may be serving a greater number of users than any other balloon, but the overall demand bandwidth supplied by this balloon is less than the other. So, considering different parameters, the priority of each node is determined and based on this, the probability can be assigned. Connectivity between the far nodes is not directly possible with radio frequency, hence the use of percolation is used for indirect connectivity or hop connectivity.

The Google Access Point which controls the balloons has two methods to do it, the nodes which are near and directly reachable, are directly controlled. The other nodes are controlled by the mesh network communication, i.e. Via other nodes by hops. For the balloon to balloon communication in a mesh network, a sophisticated protocol needs to be designed, as it involves many complexities with the dynamically changing position of the nodes (balloons are being referred to as nodes), users accessing each node and demand served by each node. In situations, where for any area, the demand increases, new nodes need to be employed. Similarly, if the needs are depreciating, nodes can be removed.

## V. PROTOCOL PROPOSAL

As the Loon Project is currently focusing on giving network connectivity to only a limited area, it can be compared to geocasting of internetwork connectivity. For connectivity between the balloons, as the radio frequency network won't be strong for too many hops, hence this paper proposes for region-based communication.

**(i) Content-Based Multicasting Routing Protocol**

In a content-based multicasting routing protocol, the region of communication is divided into blocks and each region is assigned a leader. When a node from one region wants to move into the other region, it contacts the leader of that region for the information. Similarly, this paper proposes a technique where the overall region of the balloon network can be divided into regions and each region can be assigned a leader, which may contain information about all other nodes in its group. The modification which is done in the region splitting protocol is that instead of splitting the regions into grids, it can be divided into sectors.

**(ii) Convex Hull**

As it is a wireless network, each node will not have the information of all other nodes. The main advantage of sector-wise division is that unlike grid-wise division we do not need to traverse in various directions to move from block to block, the merely clockwise or anticlockwise direction can be chosen based on the radial from the destination node.

It is known that the information of all the independent nodes is present with the Google Access Point, but the topology of the network is dynamic. Hence, the proposal introduces the concept of creating a convex hull. Even if the network topology of the balloon's changes, the convex hull will remain the same. The convex hull is a polygon formed by joining the outermost nodes of a graph, such that all the other nodes are not out of this region. The information of the convex hull is static for the balloon network even if the mesh network changes. This is being used in this proposal so that the GAP explicitly knows the region covered by the balloon network.

As the network has been stated as a dynamic network where, new balloons can be sent up and connected to the network, or any balloon can be taken off, depending on the demand and bandwidth of network connectivity of the users. When the demand for a specific region inside the convex hull increases, we can introduce a new balloon, the convex hull remains the same and this new balloon will be assigned under the leader of the sector where it lies. If the new balloon has to be put at an area outside the convex hull, then this convex hull information is discarded from the Google Access Point and taking into consideration the new balloon, the new convex hull will be formed, along with the new split in sectors and assignment of sector leaders. Similarly, if we need to remove a balloon from the network and if it lies within the convex hull, it can be easily removed, and the sector leader information must be updated. If it is a node on the hull, then a new hull must be formed, and sectors and sector leaders must be initialized.

**(iii) Dynamic Core Based Mesh Protocol**

The proposal incorporates the Dynamic Core Based Mesh protocol for inter-balloon communication. All the balloons in the mesh network will be providing the internetwork connectivity to the ground users, but not all balloons will be participating in the communication within the balloon mesh network. As mentioned earlier, the sector leaders will be containing information about all the nodes in their particular sector. In the balloon mesh network, there is no fixed source or destination, any of the balloons can be acting as a source at any point in time. It is not required for all the balloons to act as a source. As in Dynamic Core Based Mesh

Protocol, some of the nodes can be set as passive nodes that have their corresponding core active nodes, which transmit for the passive nodes.

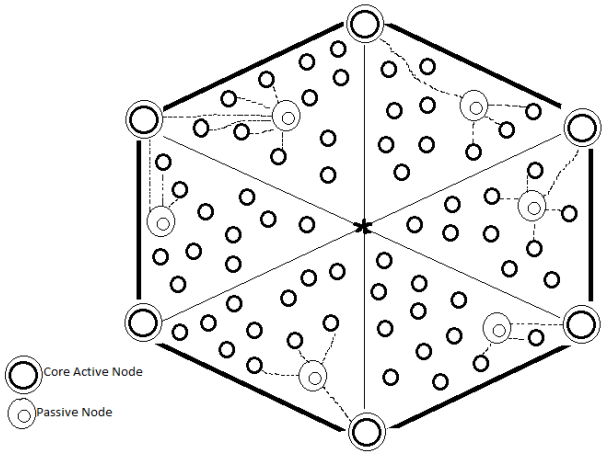

Figure 1: Convex hull and sector division of the mesh network

The sectors inside the convex hull may be divided with some angle Ө. Each sector should have a sector head. As the sectors are narrower at the center of the network area and broader at the boundaries of the hull, the sector heads are preferred to be on the outer contour of the mesh network area. The sector head may be so placed that it is within a max hop distance from the nodes forming the convex hull within its sector. This is done for the purpose that the balloons on the convex hull may be made as passive nodes and the sector heads could be acting as the core active node for them. The node forming the core-active node may also act as the leader of that sector.

The angle Ө is based on the total number of nodes in the mesh network and the number of nodes forming the convex hull. A threshold value may be fixed which defines the maximum number of nodes per sector, based on which the angle for the sectors may be adjusted when convex hull reconfiguration is done.

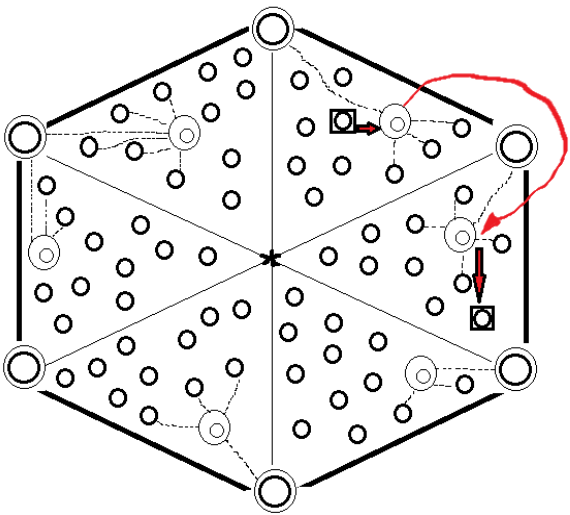

Figure 2: Transmission of data packets from source to destination based on leader-based information gathering

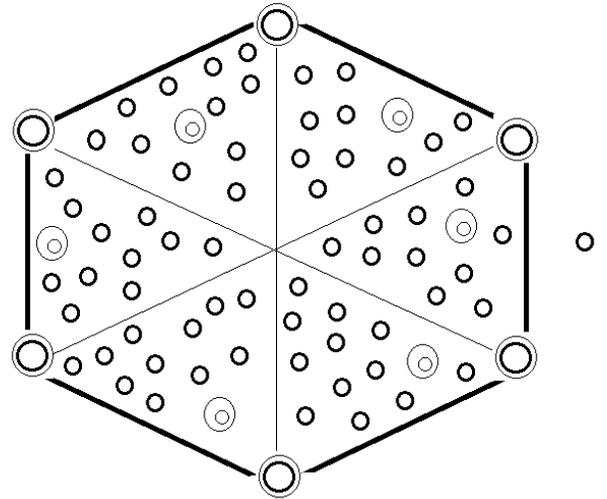

Figure 3(a): Division of the network into 4 sectors: Addition of a new node 'A' to the network/ Removal of boundary node of sector

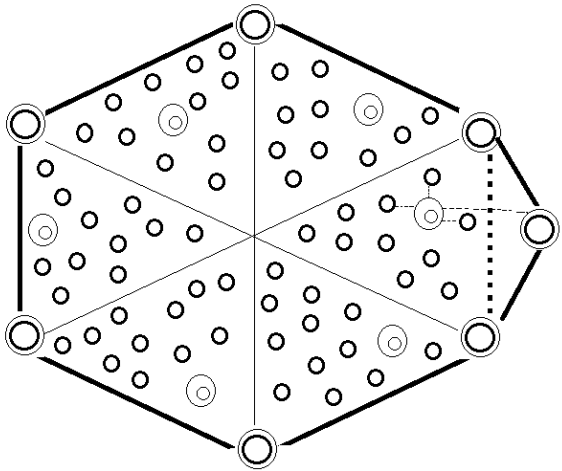

Fig 3(b): 'A' node added, mesh reconfigured

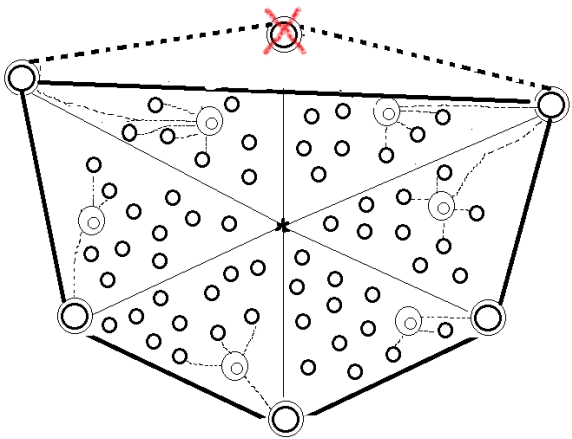

Fig 3(c): Boundary node removed, mesh reconfigured

**(iv) Advantage of sector network over a grid network**

The concept of dividing the mesh network into sectors than the grid is to decrease the parameter considerations, i.e for grid we need the x-coordinate and y-coordinate and the direction of motion can be up, down, right and left basically, whereas in sector-based implementation we just need to consider the polar coordinates of the nodes and the direction of motion is restricted to either clockwise or anticlockwise.

The reconfiguration of the partitioning of the mesh network into sectors is much easier as we know the convex hull of the network, compared to grid distribution.

In case a sector doesn't have a boundary node (convex hull node), then the leader of that sector can be selected based on the predictive estimation method considering its neighborhood sectors.

To provide better internetwork connectivity to the ground users, the balloons need to maximize their bandwidth, so they need to spend the least bandwidth for the balloon to balloon and balloon to Google Access Point communication. Hence, by using the DCB Mesh Protocol, the control overhead is reduced. Moreover, the packet delivery ratio between the balloon to balloon communication is also increased resulting in better response to the Google Access Point.

**(v) Link Probability approach for efficient transmission**

As the mesh network is formed between the nodes of the network, multicasting/ broadcasting will be taking place. There would be multiple paths between the source and the destination. The proposed protocol also takes into consideration finding the best path. This is done by the concept of random graphs, where probabilistic analysis is introduced. As mentioned earlier, each node will be having multiple parameters for operation, like bandwidth, internet range outputted, load, neighboring node details, etc. so based on the two adjacent nodes, the edge connecting both the nodes can be given a node failure probability (p) or effective transmission probability (1-p).

Apart from the other parameters, the protocol involves assigning a priority level to each node. The priority level can be defined as, Level 0- highest priority to core active/leader nodes, Level 1- medium priority to all nodes except core active and boundary nodes, Level 2- low priority to boundary/passive nodes. The priority is an important parameter as the effective link transmission probability should be high between the nodes having higher priority.

Based on the probabilities given to each node, the protocol will dynamically evaluate the best route between a specified source and destination. The protocol will dynamically model itself whenever there is a request from transmission from a source to the destination, the link having the end-to-end higher effective transmission probability will be chosen.